 \documentclass[12pt]{article}
\usepackage{enumerate}
\usepackage{amssymb}
\usepackage{amsmath}

\newcommand{\mbra}[1]{\langle {#1}|}
\newcommand{\mket}[1]{ |{#1} \rangle }

\newcommand{\hsp}{\hspace{.2 in}}

\newcommand{\beq}{\begin{eqnarray}}
\newcommand{\eeq}{\end{eqnarray}}
\linethickness{0.4pt}
\linethickness{0.4pt}
\begin{document}
\newtheorem {proposition}{Proposition}[section]
\newtheorem{lemma}{Lemma}[section] \newtheorem{theorem}{Theorem}[section]
\newtheorem{corollary}{Corollary}[section]
\begin{flushright}
\begin{center}
{\Large\bf  The exact solution of the eigenproblem for the parametric
down conversion process in the Kerr medium}
\end{center}
\end{flushright}
\begin{center}
{\bf Goce Chadzitaskos*, Maciej Horowski**, Anatol Odzijewicz**,
Agnieszka Tereszkiewicz**} \footnote{
E-mail:**chadzita@br.fjfi.cvut.cz, *aodzijew@labfiz.uwb.edu.pl,
horowski@alpha.uwb.edu.pl, tereszk@alpha.uwb.edu.pl}
\end{center}\begin{center}
\end{center}
\begin{center}
{*Faculty of Nuclear sciences and Physical Engineering \\ CTU, Brehova 7, CZ-115 19 Praha 1,
Czech Republic \\ **Institute of Theoretical Physics\\ University in Bia{\l}ystok \\Lipowa 41,
15-424 Bia{\l}ystok, Poland}
\end{center}
\bigskip
\section*{}
\begin{abstract}
The eigenproblem for a class of Hamiltonians of the parametric down conversion process in the
Kerr medium is solved. Some physical characteristics of the system are calculated.
\end{abstract}
\section{Introduction}
 During the recent years the parametric down conversion process has been intensively
investigated. It is the  simplest nonlinear model in which entangled photons (biphotons)
are produced, see \cite{P-L}. Another intensively studied phenomenon in quantum optics is
the Kerr effect, which is the simplest example of a optical bistability, \cite{P-V-L}.
In this paper we  consider  the  Hamiltonian which describes both effects simultaneously.
Following \cite{J} we chose the special form of the parametric down conversion term
and therefore our Hamiltonian is given by (\ref{11}). Using the techniques developed
in \cite{O-H-T, H-O-T}, which are strictly related to the theory of orthogonal polynomials,
we solve the eigenproblem for Hamiltonian (\ref{11}). This method works only if the parameters
 $\omega_1,\;\omega_2,\; g $ and $K_1,\; K_2$ included in the Hamiltonian, satisfy extra
conditions (\ref{a}), and then the solution is given in terms of the Dual Hahn polynomials.
     In the last section we show the way how to calculate the expecattion values of physical
quantities in  any state, in particular coherent state and occupation number state.
Physical importance of this model required to take in consideration more details, and it
will be done in a separate paper.
\section{The Hamiltonian of the system}
\renewcommand{\theequation}{2.\arabic{equation}}
\setcounter{equation}{0}
 Let $\mathbf a_1^{},\mathbf a_2^{}$ and $\mathbf a_1^*,\mathbf a_2^*$ be the usual
annihilation and creation operators of photons in the mode $1$ and $2$. They satisfy the
standard commutation relations
\beq
\label{Ab} [\mathbf a_i,\mathbf a_j^*]=\delta_{ij},
\;\;\;\;[\mathbf a_i, \mathbf a_j]=0,\;\;\;\;\;\;i,j=1,2,
\eeq
and act in the Hilbert space $\mathcal{H}$ spanned by the Fock basis
\beq
\label{Ac}
\mathcal{B}_F:=\left\{|n_1,n_2\rangle=\frac{(\mathbf a_1^*)^{n_1}(\mathbf a_2^*)^{n_2}}
{\sqrt{n_1!n_2!}}|0,0\rangle:\;n_1,n_2=0,1,\ldots\right\}
\eeq
in the usual way.    Process of quantum-optical parametric down conversion is described in
any Hamiltonian by the term  $\mathbf a_1^{2}\mathbf  a_2^{*}$,  if additionally we  need
an intensity  dependent process we must multiply this term by a function of
$\mathbf a_1^* \mathbf a_1^{}$ and $\mathbf a_2^*\mathbf a_2^{}$.
  In this paper, due to \cite{J}, we consider the system with the amplitude dependent
down conversion process described by
$ g \sqrt{\mathbf a_2^*\mathbf a_2^{}}\;\mathbf a_1^{2}\mathbf   a_2^{*} $,
where $ g $ is a real constant. The Kerr effect is described by the terms of the form
$(\mathbf a_i^*\mathbf a_i^{})^2$. Thus in order to take into consideration the free energy
of the modes we will  study the following Hamiltonian
\beq \label{11}
{\mathbf H}=\omega_1\mathbf a_1^*\mathbf a_1^{}+\omega_2 \mathbf a_2^*\mathbf a_2^{}+K_1
(\mathbf a_1^*\mathbf a_1^{})^2+K_2 (\mathbf a_2^*\mathbf a_2^{})^2
 + g  \left(\sqrt{\mathbf a_2^*\mathbf a_2^{}}\;\mathbf a_1^{2}\mathbf   a_2^{*} +
\mathbf a_1^{*2} \mathbf a_2^{} \sqrt{\mathbf a_2^*\mathbf a_2^{}}\right),
\eeq
where $\omega_1,\;\omega_2,\;K_1,\;K_2$ are real constants.
It is easy to check that the operator
\beq\label{Ad}   \mathbf{R}:= \mathbf a_1^*\mathbf a_1^{}+ 2 \mathbf a_2^*\mathbf a_2^{}
\eeq
and the projective operator $\mathbf P$ defined by
\beq   \mathbf P\mket{n_1,n_2}:=\left\{\begin{array}{cc}   0 &  \textrm{if } n_1
\textrm{ is even}\\   \mket{n_1,n_2}& \textrm{if } n_1 \textrm{ is odd}   \end{array}\right.
 \eeq
are  constants of motion, and each vector of Fock basis $  \mathcal{B}_F$ is its eigenvector.
In order to decompose the Hilbert space $\cal H$ into invariant subspaces    let us notice that
 for each nonegative integers $M$ and $ p \in\{0,1\}$  the subspace
\beq\label{Af}   \mathcal{H}_{p,M }:=\mbox{span}\left\{| 2k+ p ,M-k \rangle,\;
k=0,1,\ldots,M\right\}
\eeq
is the eigenspace of   the operators $\mathbf R $ and $\mathbf P, $ i.e.
\beq     \forall\; |\psi\rangle\in{\cal H}_{ p , M}\hsp \mathbf R \;|\psi\rangle = (2M+ p )\;
 |\psi\rangle\hsp     \mathbf P \;|\psi\rangle =  p \;  |\psi\rangle.
\eeq
     Because of this fact the original Fock space can be decomposed
\beq\label{Ag} \mathcal{H}=\bigoplus_{^{ p  \in\{0,1\}}_{M\in{\;\Bbb{N}\;}\cup\{0\}}}
\mathcal{H}_{ p , M}
\eeq
where,  of course ,  each $\mathcal{H}_{ p, M}$ is $\mathbf H$-invariant and of finite dimension
\beq\label{Agg}
 \dim  \mathcal{H}_{ p, M}=M+1.
 \eeq
The eigenproblem for the Hamiltonian (\ref{11}) is reduced to the eigenproblems of the
finite--dimensional  operators $\mathbf{H}_{ p , M}:=\mathbf{H}|_ {\mathcal{H}_{ p , M}}$.
The matrix form of $\mathbf{H}_{ p, M}$ in the basis
$\left\{| 2k+ p ,M-k \rangle \right\}_{k=0}^M$ is
\begin{equation}\label{Al}
\mathbf{H}_{ p, M} = 2  g  \cdot\begin{pmatrix}   a_{0} & b_{0} & 0     &\cdots & 0 \\
 b_{0} & a_{1} & b_{1} &       & \vdots \\
  0    & b_1   &a_2    &\ddots &  0\\
 \vdots &       & \ddots&\ddots &b_{M-1} \\
   0 & \cdots    & 0     &b_{M-1}&a_M \\
 \end{pmatrix}
\;\;\;+\;C_{ p , M }\cdot
\begin{pmatrix}  1& 0 & 0     &\cdots & 0 \\ 0 &1 & 0 &       & \vdots \\
0    & 0   &1   &\ddots &  0\\  \vdots &       & \ddots&\ddots &0 \\
 0 & \cdots    & 0     &0&1 \\
\end{pmatrix},
\end{equation}
where
\beq\label{Am}   a_k&=&\frac{k^2}{2 g }(4K_1+K_2)+\frac{k}{2 g }(2\omega_1-\omega_2+4 p
K_1-2MK_2)+M( p +\frac{1}{2}),\\      b_k&=&(M-k)\sqrt{(k+1)(k+ p +\frac{1}{2})} ,
\nonumber
\eeq       and the constant is
\begin{equation}\label{An} C_{ p ,M }=-2 g  M\left( p +\frac{1}{2}\right)+\omega_1 p +
\omega_2 M+K_1 p +K_2M^2.
\end{equation}
In the next section we show that under some special conditions for parameters
$\omega_1,$ $\omega_2,$ $ g ,$  $K_1,$ $K_2$ the first matrix of  (\ref{Al}) can be
diagonalized  by the Dual Hahn polynomials theory.
\section{Dual Hahn polynomials and spectrum of the Hamiltonian   }
\renewcommand{\theequation}{3.\arabic{equation}}
\setcounter{equation}{0}
We start this section with the brief definition and some important properties of
the Dual Hahn polynomials \cite{K-S}.  For any fixed natural number $N$ and for
$\gamma>-1$, $\delta>-1$ let us define a finite set of real points
\beq\label{A1}
 \lambda_l=l(l+\gamma+\delta+1),\;\;l=0,1,\ldots,N\;.
 \eeq
The Dual Hahn orthonormal polynomials are, by definition, the finite family of
the polynomials  $\left\{P_k(\lambda_l;\gamma,\delta,N) \right\}_{k=0}^N$
of the discrete variable $\lambda_l$ satisfying the following three term recurrence formula
\beq\label{A2}
\lambda_l P_k(\lambda_l;\gamma,\delta,N)=b_kP_{k+1}(\lambda_l;\gamma,\delta,N)
+a_k P_k(\lambda_l;\gamma,\delta,N)+ b_{k-1}P_{k-1}(\lambda_l;\gamma,\delta,N),
\eeq
where $a_k,\;b_k$ are defined by
\beq
a_k&=&(k+\gamma+1)(N-k)+k(N+\delta+1-k)\nonumber\\\label{3}\\
b_k&=&\sqrt{(k+1)(k+\gamma+1)(N-k)(N+\delta-k)}.\nonumber
\eeq
Formula (\ref{A2}) must be satisfied for any $k,l=0,1,\ldots,N$.
The orthonormality means
 \beq\label{A5}
\sum\limits_{l=0}^{N}  \varrho_{ N}(l;\gamma,\delta)P_r(\lambda_l;\gamma,\delta,N)P_s(\lambda_l;\gamma,\delta,N)
=\delta_{rs},
  \eeq
where
\beq\label{A4}
\varrho_{ N}(l;\gamma,\delta)=\frac{(2l+\gamma+\delta+1)(\gamma+1)_l(-N)_l\;N!}
{(-1)^l(l+\gamma+\delta+1)_{N+1} (\delta+1)_l\;l!},
\eeq
and the Pochhammer-symbol $(a)_l$ is defined by the gamma  function
\beq\label{A6}
  (a)_l:=\frac{\Gamma(a+l)}{\Gamma(a)}.
\eeq
Let us note that the Dual Hahn orthonormal polynomials can be expressed in terms of the
hypergeometric functions
\beq\label{A7}
P_k(\lambda_l;\gamma,\delta,N)=\sqrt{\frac{\Gamma(\gamma+k+1)\Gamma(\delta+N-k+1)}
    {k!(N-k)!\Gamma(\gamma+1)\Gamma(\delta+1)}}\;\;_3F_2\left(^{-k,\;-l,\;l+\gamma+\delta+1}_{\;\;\;\;\;
          \gamma+1,\;-N}\mid 1\right).
\eeq Now we use this polynomials to solve the eigenproblem of  the
Hamiltonian (\ref{Al}) for the case when $\omega_1,$ $\omega_2,$ $
g ,$  $K_1,$ $K_2$ obey special conditions \beq   \nonumber
2\omega_1-\omega_2- g =0\label{a}\\ 2K_1+ g =0\\ K_2+2 g =0.
\nonumber \eeq We observe that the substitution \beq   \gamma = p
- \frac{1}{2} \hsp \delta=0 \hsp N=M\label{Ao} \eeq realizes the
equivalence between the formulae (\ref{Am}) and (\ref{3}).  Hence
the eigenproblem
\beq\label{Ar}
{\mathbf H}_{ p , M}|E_{l, p ,M}\rangle =E_{l, p ,M}|E_{l, p
,M}\rangle \hsp l=0,1,\ldots M \eeq
 due to (\ref{A2})
 is solved  in terms of Dual Hahn polynomials,
 \beq\label{As}
E_{l, p ,M}&=&2 g \;\lambda_l  -2 g  M\left( p +\frac{1}{2}\right)+
\omega_1 p +\omega_2 M+K_1 p ^2+K_2M^2\\   |E_{l, p ,M}\rangle&=&\label{At}
  \sqrt{ \varrho_{ M}(l; p -\frac{1}{2},0 )}  \sum\limits_{k=0}^M P_k(\lambda_l;
p -\frac{1}{2},0 ; M)\;\; | 2k+ p ,M-k \rangle.
\eeq
 The multiplier on the right hand side of the last formula is chosen to fulfil the
orthonormality condition
\beq\label{Au}
\langle E_{k, p ,M}|E_{k', p ',M'}\rangle =\delta_{kk'}\;\delta_{ p  p '}\;\delta_{MM'},
 \eeq           what follows from (\ref{A5}).
We can also express the vectors $| 2k+ p ,M-k \rangle$ in the
basis of the eigenvectors of ${\mathbf H}_{ p ,M}$ : \beq  | 2k+ p
,M-k \rangle&:=&\label{Aw}  \sum\limits_{l=0}^M \sqrt{ \varrho_{
M} (l; p -\frac{1}{2},0 )   }P_k(\lambda_l; p -\frac{1}{2},0
;M)\;\; |E_{l, p ,M}\rangle.
 \eeq
\section{Physical characteristics of the system}
\renewcommand{\theequation}{4.\arabic{equation}}
\setcounter{equation}{0}
In this section we describe the time evolution of expectation values of physically
interesting observable on arbitrary state $\mket{\psi}$. To do this, it is important
to calculate the matrix elements of the evolution operator $e^{-i\mathbf Ht}$.
Because ${\mathcal H}_{ p ,M}$ is $\mathbf H$ invariant, the only nonvanishing matrix
elements of $e^{-i \mathbf H t}$ are of the form
\beq
\label{Bccc} \langle \; 2l+ p ,M-l\;|e^{-i\mathbf Ht}\;|2k+ p ,M-k \; \rangle &=
&\\ &&\!\!\!\!\!\!\!\!\!\!\!\!\!\!\!\!\!\!\!\! \!\!\!\!\!\!\!\!\!\!\!\!\!\!\!\!\!\!\!\!\!\!
\!\!\!\!\!\!\!\!\!\!\!\!\!\!\!\!\!\! \!\!\!\!\!\!\!\!\!\!\!\!\!\!\!\!\!\!\!\!\!\!\!\!\!\!\!\!
\!\!\!\!\!\!\!\!\!\!\!\!\!\!
\nonumber=\sum_{j=0}^{M}  e^{-i\;t\;E_{j, p ,M}}\; \varrho_{M}(j; p -\frac{1}{2},0 )
P_l(\lambda_j; p -\frac{1}{2},0 , M) P_k(\lambda_j; p -\frac{1}{2},0  ,M)
 \eeq
what follows from (\ref{Aw}) and (\ref{Ar}). The decomposition (\ref{Ag}) leads us to the
following resolution of identity operator $\mathbf 1 $ in ${\mathcal H}$
\beq \label{Bcc} \mathbf 1=\sum_{M=0}^{\infty}    \sum_{k=0}^{M}\sum_{ p =0}^1\;|2k+ p ,M-k\rangle
\langle 2k+ p ,M-k\,|.
\eeq
Let $\mathbf X$ be any observable and $|\psi\rangle$ any arbitrary chosen initial state.
Then using four times the resolution (\ref{Bcc}) and (\ref{Bccc}) we obtain that the time
evolution of the expectation value of $\mathbf X$ in the state $|\psi\rangle$ is given by
\beq\label{z}
\langle\psi|e^{i{\mathbf H}t}\mathbf X   e^{-i{\mathbf H}t}|\psi\rangle  & =&
\sum_{M_1,M_4=0}^{\infty} \sum_{ p _1, p _4 =0}^{1}
\sum_{k_1,k_2=0}^{M_1}\sum_{k_3,k_4=0}^{M_4}  \langle \psi\;|2k_1+ p _1,M_1-k_1\rangle \times\\
 &&
\times\langle \;2k_1+ p _1,M_1-k_1\;|e^{i\mathbf Ht}\;|2k_2+ p
_1,M_1-k_2 \; \rangle \;\times\nonumber\\
 &&
 \times \langle 2k_2+ p _1,M_1-k_2|\;\mathbf X\;
   2k_3+ p _4,M_4-k_3\rangle\;\times\;\nonumber\\&&\times\langle \;
  2k_3+ p _4,M_4-k_3\;|e^{-i\mathbf Ht}
|k_4+ p _4,M_4-k_4 \; \rangle\;\langle k_4+ p _4,M_4-k_4      |\;\psi\rangle\nonumber.
 \eeq
Substituting $\mket{\psi}=\mket{2l+ p ,M-l} $ in (\ref{z}) and using (\ref{Bccc})
we obtain as a finite sum
\beq
\mbra{2l+ p ,M-l}e^{i{\mathbf H}t}\mathbf X
e^{-i{\mathbf H}t}\mket{2l+ p ,M-l}    &=&\\
&& \! \! \! \! \! \! \! \! \! \! \! \! \! \! \! \! \! \! \! \! \! \! \! \! \! \! \! \!
  \! \! \! \! \! \! \! \! \! \!    \! \! \! \! \! \! \! \! \! \! \! \! \! \! \! \! \! \! \!
  \! \! \! \! \! \! \! \! \! \! \! \! \! \! \! \! \! \! \! \! \! \! \! \! \! \! \! \! \! \! \!
\! \! \! \! \! \! \!    \! \! \! \! \! \! \! \! \! \! \! \! \! \! \! \! \! \!    =
  \sum_{j_1,j_2,k_2,k_3=0}^{M}            e^{-2\;i\;t\; g (\lambda_{j_2}-\lambda_{j_1})}\;
 P_l(\lambda_{j_1}; p -\frac{1}{2},0 , M) P_{k_2}(\lambda_{j_1}; p -\frac{1}{2},0  ,M)
  \times\nonumber\\   && \! \! \! \! \! \! \! \! \! \! \! \! \! \! \! \! \! \! \! \! \! \! \!
\! \! \! \! \!  \! \! \! \! \! \! \! \! \! \!    \! \! \! \! \! \! \! \! \! \! \! \! \! \! \!
\! \! \! \!    \! \! \! \! \! \! \! \! \! \! \! \! \! \! \! \! \! \! \! \! \! \! \! \! \!
\! \! \! \! \! \! \! \! \! \! \! \! \!    \! \! \! \! \! \! \! \! \! \! \! \! \! \! \! \!
\! \!    \times    \varrho_{M}(j_1; p -\frac{1}{2},0 )\;
\varrho_{M}(j_2; p -\frac{1}{2},0 ) P_{k_3}(\lambda_{j_2}; p -\frac{1}{2},0  ,M)
P_{l}(\lambda_{j_2}; p -\frac{1}{2},0  ,M)   \times\nonumber\\
&& \! \! \! \! \! \! \! \! \! \! \! \! \! \! \! \! \! \! \! \! \! \! \! \! \! \! \! \!
\! \! \! \! \! \! \! \! \! \!    \! \! \! \! \! \! \! \! \! \! \! \! \! \! \! \! \! \! \!
\! \! \! \! \! \! \! \! \! \! \! \! \! \! \! \! \! \! \!\! \! \! \! \! \! \! \! \! \! \!
\! \! \! \! \! \!\nonumber
\times\mbra{2k_2+ p ,M-k_2} \;\mathbf X\;   \mket{2k_3+ p ,M-  k_3}.
\eeq
Moreover, when we put for example $\mathbf X(t)=\mathbf a_1^*(t)\mathbf a_1^{}(t),$ then
\beq     \mbra{2l+ p ,M-l}e^{i{\mathbf H}t}\mathbf a_1^*(t)\mathbf a_1^{}(t)
e^{-i{\mathbf H}t}\mket{2l+ p ,M-l}    &=&\\
&& \! \! \! \! \! \! \! \! \! \! \! \! \! \! \! \! \! \! \! \! \! \! \! \! \! \! \! \!
\! \! \! \! \! \! \! \! \! \! \! \! \! \! \! \! \! \! \! \! \! \! \! \! \! \! \! \! \!
 \! \! \! \! \! \! \! \! \! \! \! \! \! \! \! \! \! \! \! \! \! \! \! \! \! \! \! \! \!
\! \! \! \! \! \! \! \! \!    \! \! \! \! \! \! \! \! \! \! \! \! \! \! \! \! \! \!    =
p +2     \sum_{j_1,j_2,k=0}^{M}         k\;   e^{-2\;i\;t\; g (\lambda_{j_2}-\lambda_{j_1})}\;
 P_l(\lambda_{j_1}; p -\frac{1}{2},0 , M) P_{k}(\lambda_{j_1}; p -\frac{1}{2},0  ,M)
 \times\nonumber\\
&& \! \! \! \! \! \! \! \! \! \! \! \! \! \! \! \! \! \! \! \! \! \! \! \! \! \! \! \!
\! \! \! \! \! \! \! \! \! \!    \! \! \! \! \! \! \! \! \! \! \! \! \! \! \! \! \! \! \!
   \! \! \! \! \! \! \! \! \! \! \! \! \! \! \! \! \! \! \! \! \! \! \! \! \! \! \! \! \! \!
\! \! \! \! \! \! \! \!    \! \! \! \! \! \! \! \! \! \! \! \! \! \! \! \! \! \!
  \times   \varrho_{M}(j_1; p -\frac{1}{2},0 )\;    \varrho_{M}(j_2; p -\frac{1}{2},0 )
P_{k}(\lambda_{j_2}; p -\frac{1}{2},0  ,M)     P_{l}(\lambda_{j_2}; p -\frac{1}{2},0  ,M)
\nonumber.
\eeq
Now the calculation of the dispersion for this operator is straightforward, too.
When we consider a coherent state
$\mket{\psi}=\mket{z_1,z_2}=\sum_{n_1,n_2=0}^\infty
\frac{z_1^{n_1}z_2^{n_2}}{\sqrt{n_1!n_2!}}\mket{n_1,n_2}$
then (\ref{z}) takes the form
\beq\label{aa}
&& \mbra{z_1,z_2}e^{i{\mathbf H}t}\mathbf X   e^{-i{\mathbf H}t}\mket{z_1,z_2}=\\  &&
\sum_{M_1,M_4=0}^{\infty} \sum_{ p _1, p _4 =0}^{1}    \sum_{k_1,k_2=0}^{M_1}
\sum_{k_3,k_4=0}^{M_4} \frac{z_1^{ p _1+ p _4}z_2^{M_1+M_4}}{\sqrt{(2k_1+ p _1)!(2k_4+ p _4)!
(M_1-k_1)!(M_4-k_4)!}} \left(\frac{z_1^2}{z_2}\right)^{k_1+k_4}\nonumber\times\\
&&\times\langle \; 2k_1+ p _1, M_1-k_1\;|e^{i\mathbf Ht}\;|2k_2+ p _1, M_1-k_2\; \rangle\nonumber
\times\\ &&\times\langle 2k_2+ p _1 ,M_1-k_2|\;\mathbf X\;
  2k_3+ p _4,M_4-k_3\rangle\langle \;2 k_3+ p _4,M_4-k_3\;|e^{-i\mathbf Ht}\;|2k_4+
p _4,M_4-k_4 \;            \rangle\nonumber.
\eeq
The formulas (\ref{z})-(\ref{aa}) initialize a new area of investigations and numeric
calculations. It is another thing to study this ideas according to the spirit
of  the theory presented in \cite{H-O-T}.
This short paper do not allow us to expand this subject. Because of the physical
importance of this theme we will devote to this subject a separate paper.
\section*{ Acknowledgements}
We would like to thanks J.Tolar and T. Goliñski for them interest
in the subject. The authors wish to acknowledge partial support of
the Ministry of Education of Czech Republic under the research
project MSM210000018.


\begin{thebibliography}{99}
\bibitem [P-L]{P-L} V. Pe\v{r}inov\'{a}, A.Luk\v{s} "Parametric down-conversion experiments
with stationary fields", Forstchr. Phys.\textbf{51}, No.2-3, 211-218 (2003)
\bibitem [P-V-L]{P-V-L} V. Pe\v{r}inov\'{a}, V. Vrana A.Luk\v{s}, "Quantum statistics of
displaced Kerr states", Phys. Rev. A,\textbf{51}, No. 3, 2499-2514 (1995)
\bibitem [H-O-T]{H-O-T}   M. Horowski, A.Odzijewicz and A.Tereszkiewicz, "Some integrable
system    in nonlinear quantum optics",  J.Math.Phys. 44, (2003) 2, (480-506),
arXiv:math-ph/0207031 v1 23 Jul 2002
\bibitem [O-H-T]{O-H-T} A. Odzijewicz, M. Horowski, A.Tereszkiewicz "Integrable multi-boson
systems and orthogonal polynomials" J. Phys. A: Math. Gen. 34, (2001) 4353-4376
\bibitem [K-S]{K-S} R. Koekoek, R.F.Swarttouw, "The Askey-scheme of hypergeometric orthogonal
polynomials and its q-analoue",http://aw.twi.tudelf.nl
\bibitem [J]{J} I.Jex, G. Drobny, Phys. Rew. A, 47, 3251 (1993)
\end{thebibliography}
 \end{document}